\pdfoutput=1
\documentclass[11pt]{article}

\usepackage[final]{acl}

\usepackage{times}
\usepackage{latexsym}
\usepackage{textcomp}
\usepackage[T1]{fontenc}
\usepackage[utf8]{inputenc}
\usepackage{microtype}
\usepackage{inconsolata}
\usepackage{booktabs}
\usepackage{xspace}
\usepackage{amsmath}
\usepackage{amssymb}
\usepackage{mathtools}
\usepackage{tcolorbox}
\usepackage{xspace}
\usepackage{caption}
\usepackage{subcaption}
\usepackage{balance}
\usepackage[hang,flushmargin]{footmisc}
\usepackage{multirow}

\newcommand{\ourmodel}{VISA}

\title{VISA: Retrieval Augmented Generation with  Visual Source Attribution}

\author{
Xueguang Ma\thanks{\quad Equal contribution}$^1$\quad
Shengyao Zhuang$^*$$^{2,3}$\quad
\textbf{Bevan Koopman}$^{2,3}$\quad\\
\textbf{Guido Zuccon}$^3$\quad
\textbf{Wenhu Chen}$^1$\quad
\textbf{Jimmy Lin}$^1$\\
[1ex]
 $^1$University of Waterloo \quad $^2$CSIRO \quad $^3$ University of Queensland
}
\begin{document}
\maketitle
\begin{abstract}
Generation with source attribution is important for enhancing the verifiability of retrieval-augmented generation (RAG) systems.
However, existing approaches in RAG primarily link generated content to document-level references, making it challenging for users to locate evidence among multiple content-rich retrieved documents.
To address this challenge, we propose \textit{Retrieval-Augmented Generation with \underline{Vi}sual \underline{S}ource \underline{A}ttribution} (\ourmodel), a novel approach that combines answer generation with 
visual source attribution.
Leveraging large vision-language models (VLMs), \ourmodel{} identifies the evidence and highlights the exact regions that support the generated answers with bounding boxes in the retrieved document screenshots.
To evaluate its effectiveness, we curated two datasets: Wiki-\ourmodel{}, based on crawled Wikipedia webpage screenshots, and Paper-\ourmodel{}, derived from PubLayNet and tailored to the medical domain.
Experimental results demonstrate the effectiveness of \ourmodel{} for visual source attribution on documents' original look, as well as highlighting the challenges for improvement.
Code, data, and model checkpoints will be released.

\end{abstract}

\section{Introduction}

\begin{figure*}
    \centering
    \includegraphics[width=\textwidth]{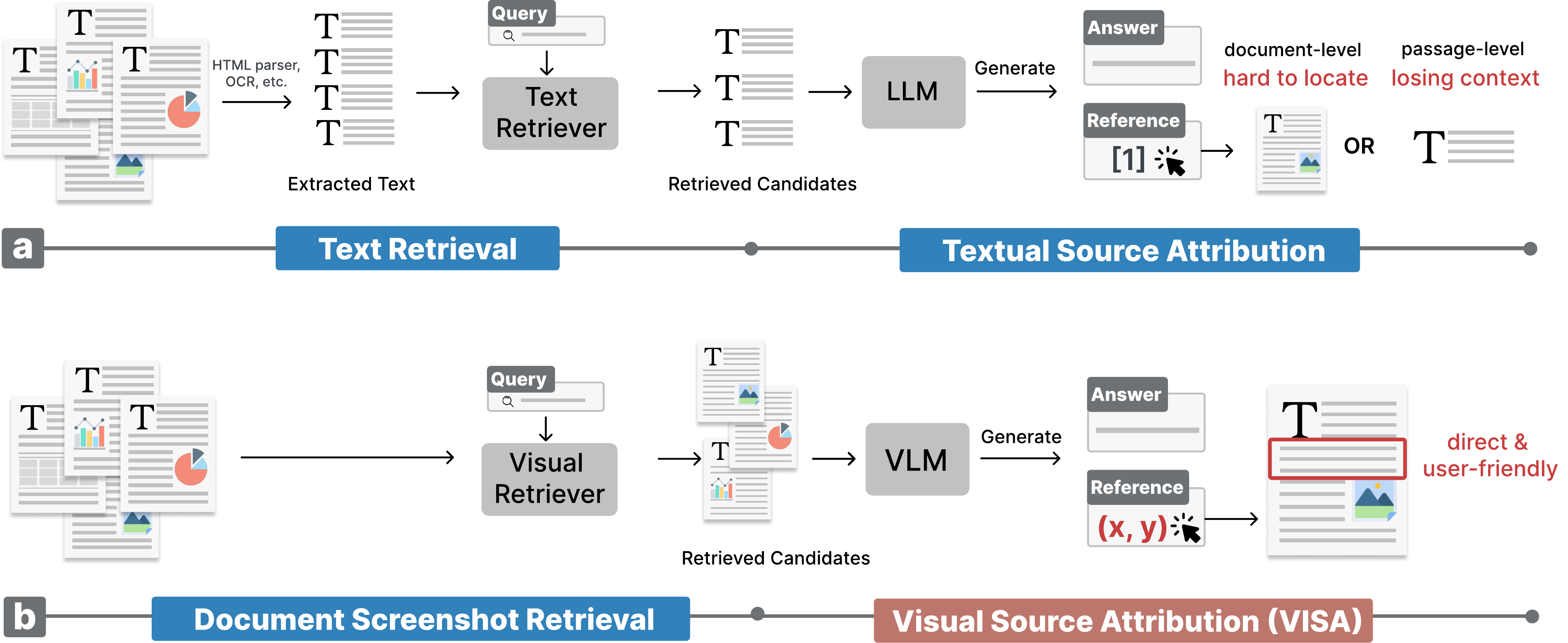}
    \caption{Comparison between (a) Text-based generation with source attribution in a RAG pipeline. and (b) Visual-based generation with source attribution in a V-RAG pipeline.
    \ourmodel{} directly pinpoint the source evidence of the answer for user query in the original document with a bounding box.}
    \label{fig:paradigm}

\end{figure*}

Retrieval-augmented generation (RAG) has become a key technique for enhancing the reliability in information-seeking processes~\cite{gao2024retrievalaugmentedgenerationlargelanguage}.
Traditional RAG pipeline directly generates an answer to a user query from retrieved candidate documents~\cite{chen2017reading, rag}.
Yet, it is hard for users to verify the sources and appropriately trust generated answers, given that models could produce hallucinated content~\cite{min-etal-2023-factscore, malaviya-etal-2024-expertqa}.
Recent works have introduced the generation with citation paradigm~\cite{gao-etal-2023-enabling, ye-etal-2024-effective},
prompting the model to not only generate answers but also directly cite the identifiers of the source documents. 
Such source attribution approaches make it possible for users to check the reliability of the outputs~\cite{asai2024reliableadaptableattributablelanguage}.
% These source attribution methods enable users to trace the generated information back to its source document, allowing them to verify the answer or explore additional knowledge from referenced documents~\cite{asai2024reliableadaptableattributablelanguage}.

However, text-based generation with source attribution faces several issues:
First, citing the source at the document level could impose a heavy cognitive burden on users~\cite{Foster1979,SWELLER201137}, where users often struggle to locate the core evidence at the section or passage level within the dense and multi-page document. 
Despite such granularity mismatch could be addressed through passage-citation-based generation methods --- linking answers to specific text chunks, it requires non-trivial extra engineering efforts to match the chunk in the document source.
Moreover, visually highlighting text chunks in the source document is more intuitive for users, but it remains challenging as it requires control over document rendering, which is not always accessible, such as in PDF scenarios.

Inspired by the recent document screenshot embedding retrieval paradigm --- dropping the document processing module and directly using VLM to preserve the content integrity and encoding document screenshots for retrieval~\cite{ma-etal-2024-unifying}, we ask whether source attribution can also be integrated into such a unified visual paradigm to establish a fully visual, end-to-end verifiable RAG pipeline that is both user-friendly and effective?

To this end, we propose \textit{Retrieval Augmented Generation with \underline{Vi}sual \underline{S}ource \underline{A}ttribution} (\ourmodel).
In our approach, a large vision-language model (VLM) processes single or multiple retrieved document images and not only generates an answer to the user query but also returns the bounding box of the relevant region within the evidence document.
As illustrated in Figure~\ref{fig:paradigm}, this method enables direct attribution by visually pinpointing the exact position within the document, allowing users to quickly check the supporting evidence within the original context for the generated answer.
VLMs are not restricted by document format or rendering, making them more versatile for diverse use cases.
Moreover, this task serves as a meaningful evaluation of VLMs, assessing their ability to provide self-explanations and accurately localize supporting information within their original input in an RAG paradigm.
To the best of our knowledge, this is the first work to leverage a VLM for directly enabling visual source attribution in an RAG framework.

To train and evaluate \ourmodel{}, we curated two datasets: Wiki-\ourmodel{} and Paper-\ourmodel{}.
Wiki-\ourmodel{} is derived from the Natural Questions dataset~\cite{kwiatkowski-etal-2019-natural}.
It reconstructs the original Wikipedia webpages, using short answers as generation targets and corresponding long answer's HTML bounding box as source attribution targets.
This dataset supports the test of model's ability to attribute sources across multi-document, multi-page, and multi-modal content.
On the other hand, Paper-\ourmodel{}, built from PubLayNet~\cite{publaynet} with synthetic query generation, focuses on the biomedical domain by evaluating performance on multi-modal scientific paper PDFs.
Together, they provide diverse and challenging benchmarks for assessing the granularity and accuracy of source attribution in RAG systems.
Our experiments, spanning both in-domain training and zero-shot evaluation, revealed existing state-of-the-art models like QWen2-VL-72B~\cite{wang2024qwen2vlenhancingvisionlanguagemodels} struggle with precise visual source attribution in zero-shot prompting.
Fine-tuning \ourmodel{} on our curated datasets significantly improved model performance in visual attribution accuracy.
Further analysis highlights key areas for improvement, such as enhancing bounding box precision for long image documents, multi-documents, and zero-shot generalization capabilities.

\section{Related Work}
\subsection{RAG attribution}
Open-domain question answering with LLMs often suffer from two key issues: hallucinations and outdated internal knowledge.
Retrieval-Augmented Generation (RAG) has been recognized as an effective solution to these problems~\cite{rag,gao2024retrievalaugmentedgenerationlargelanguage,ovadia-etal-2024-fine}.
In RAG, relevant documents are first retrieved from an external database and then fed into LLMs alongside the question.
This allows LLMs to reference the retrieved documents during answer generation.
Furthermore, RAG can generate a list of citations attached to the generated answers, linking them to the retrieved documents so users can verify the accuracy of the output.
This process is known as source attribution~\cite{rashkin-etal-2023-measuring,bohnet2023attributedquestionansweringevaluation,khalifa2024sourceaware}.

Typically, RAG with source attribution follows a text-only pipeline where all inputs and outputs, such as questions, retrieved documents, generated answers, and citations, are in textual form.
Recently, vision-based RAG pipelines have emerged, where the retrieved documents are represented as screenshot images~\cite{ma-etal-2024-unifying,faysse2024colpaliefficientdocumentretrieval}, and VLMs process both textual questions and these document images to generate answers~\cite{riedler2024textoptimizingragmultimodal,xia2024mmedragversatilemultimodalrag,yu2024visragvisionbasedretrievalaugmentedgeneration,cho2024m3docragmultimodalretrievalneed}.
Compared to traditional text-only RAG, vision-based RAG can leverage structured and visual information from documents, such as tables, graphs, and images, which are often challenging to extract through text-only pipelines. 

Our VISA attribution method proposed in this paper is a novel approach for vision-based RAG pipelines: directly drawing bounding boxes around the content in retrieved document screenshots that potentially supports the generated answers. 
This approach differs from existing attribution methods in two ways:
(1) Granularity: Existing attribution methods often operate at the document level, requiring users to read entire documents to locate supportive content. In contrast, our method directly attributes the answer to specific content within the document, such as a passage, table, or image in the screenshot.
(2) Presentation: Traditional attribution methods provide a list of textual citations, whereas our method uses bounding boxes, offering a visually-oriented form of attribution. This can help users quickly locate the relevant information.

\subsection{Bounding Box Drawing with VLM}
Bounding box-based object detection is a well-established task in computer vision (CV)~\cite{zhao2019objectdetectiondeeplearning,zou2023objectdetection20years}.
Traditional approaches rely on convolutional neural networks (CNNs)~\cite{lecun2015deep} or Vision Transformers (ViTs)~\cite{dosovitskiy2021an} to extract features and predict bounding boxes alongside object classification~\cite{NIPS2015_14bfa6bb,NIPS2016_577ef115,YOLO,DETR}. 

Recent vision-language models (VLMs) like GPT4o~\cite{openai2024gpt4ocard}, QWen2-VL~\cite{wang2024qwen2vlenhancingvisionlanguagemodels}, and PaliGemma~\cite{steiner2024paligemma2familyversatile} have shown the ability to generate bounding box coordinates in an image-to-text manner, taking input images and generate the top-left and bottom-right coordinates of target objects.
Unlike traditional object detection that focuses on natural images, our method applies bounding box drawing to text-intensive document screenshots.

Additionally, grounding elements on screenshots has been explored in GUI agent systems~\cite{cheng-etal-2024-seeclick,lin2024training}, where bounding boxes are used to localize UI elements like buttons. While these approaches focus on GUI contexts, our work targets visual source attribution in vision-based RAG processes, grounding bounding boxes to locate evidence within document images.

\section{Method}
\subsection{Task Definition}
Our \ourmodel{} is a novel source attribution method primarily designed for vision-based RAG systems.
To formally define the task of RAG with visual-based source attribution: given a textual user query $q$ as the RAG system input, the retrieval component of the system needs to retrieve a set of candidate documents $ D=\{d_1, ..., d_n\}$ from corpus $\mathcal{C}$.
Then the generation component of the system needs to return three outputs: an answer $a$ that answers the query $q$, the identifier $i$ of the most relevant document $d^*$ in $D$, and a bounding box coordinates $B_{d^*}=[(x_1, y_1), (x_2, y_2)]$ within $d_*$ that highlight the content supporting the generated answer $a$.

In a vision-based RAG setup, user queries are textual, while all documents in the corpus $\mathcal{C}$ are screenshots of documents (e.g., webpages or PDF pages) provided as image inputs.
% Consequently, both the retrieval and generation components of the RAG system must handle such multimodal inputs effectively.

\subsection{Generation with Visual Source Attribution}
This paper focuses on \ourmodel{} within the generation component of vision-based RAG systems. As discussed in the previous section, \ourmodel{} must handle multimodal input. To achieve this, we leverage VLMs for implementing \ourmodel{}. Specifically, for a given query and a set of retrieved candidate documents (i.e., screenshots of documents), the system processes the inputs as follows: query tokens are directly input into the language model, while document screenshots are first processed by the image encoder to extract image representations, which are then fed into the language model.

The language model subsequently generates the answer, the identifier of the relevant document, and the xy-coordinates of the bounding box's top-left and bottom-right corner on the content that supports the generated answer.
Notably, this entire process can be framed as a next-token prediction task.
Finally, the generated identifier and bounding box coordinates are used to draw the bounding box on the target document screenshot, which is presented to the user along with the generated answer.

Technically, existing instruction-tuned VLMs, such as Qwen2-VL-72B~\cite{wang2024qwen2vlenhancingvisionlanguagemodels}, can potentially be prompted to perform \ourmodel{} in a zero-shot manner.
However, we find that \ourmodel{} remains a challenging task.
Consequently, further supervised fine-tuning on a dedicated \ourmodel{} task dataset is necessary.
In the next section, we introduce the datasets we crafted specifically for training and evaluating \ourmodel{}.

\subsection{Dataset Acquisition}
\label{sec:dataset}
The training and evaluation data suitable for the \ourmodel{} task needs to be formatted as follows: 
the input consists of a textual query and document screenshot images as multimodal inputs, 
while the target outputs include the textual short answer, the relevant document identifier, and the coordinates of the bounding box.
To create datasets that meet these requirements, we craft existing publicly available datasets to support the training and evaluation of our proposed \ourmodel{} method.

\textbf{Wiki-\ourmodel{}} is derived from the Natural Questions (NQ) dataset~\cite{kwiatkowski-etal-2019-natural}.
The original NQ dataset provides natural questions, along with short and long answers sourced from Wikipedia webpages.
We use the short answers as answer targets.
However, the original dataset does not contain the original webpage screenshots.
We use the Selenium Python toolkit\footnote{\url{https://pypi.org/project/selenium/}} to access and render the webpage with the original URL with a history version stamp.
And take a screenshot with 980 pixels width and up to 3920 pixels (4 pages) height.
Using the long answer, we identify the corresponding element in the HTML from which the long answer is derived.
We then draw a bounding box around this element to obtain thecoordinates.
Notably, the answers in this dataset can come from various elements, such as passages, tables, lists, or images within the webpage.
Since the questions and answers in Wiki-\ourmodel{} are human-judged, we consider this dataset a high-quality, supervised dataset and evaluation for \ourmodel{} on general knowledge, with Wikipedia webpage.

\textbf{Paper-\ourmodel{}} is derived from PubLayNet~\cite{publaynet}, a dataset originally designed for document layout analysis of single page PubMed PDF documents.
PubLayNet provides bounding box coordinates and class labels (e.g., title, text, table, figure, etc.) for each element in a paper’s PDF screenshot.
However, the dataset does not include queries or answers associated with each document.
To address this limitation, we leverage instruction-tuned VLMs (e.g. Qwen2-VL-72B) to synthetically generate queries and answers. Specifically, for each paper screenshot sample in the PubLayNet training data, we select a bounding box within the sample and overlay it on the screenshot.
The modified screenshot is then input to the VLM with a prompt designed to instruct the model to generate a question and a short answer based on the content within the bounding box.
See Appendix~\ref{sec:synthetic_prompt} for the prompt details and generation example.
By augmenting the original PubLayNet in this way, we create synthetic queries and answers, enabling it to support \ourmodel{} training.
We consider the resulting Paper-\ourmodel{} dataset as synthetic training and evaluation for scientific paper PDFs in the medical domain.
% \red{add some clarification about synthetic if necessary}

\textbf{FineWeb-\ourmodel{}} is based on the FineWeb-edu corpus~\cite{penedo2024finewebdatasetsdecantingweb}, a high-quality text corpus of crawled webpages.
We sampled 60k webpage URLs and used Selenium to capture screenshots of diverse, content-rich webpages.
A passage containing more than 50 words was randomly selected as the target source.
A bounding box was drawn around the selected content, and a VLM was prompted to generate a query and short answer supported by the target content, similar as Paper-\ourmodel{}.
Although Fineweb-\ourmodel{} provides diverse layout, it do not guaranteed to high quality data has human annotated in Wiki-\ourmodel{} or Paper-\ourmodel{} that assessing a specific domain, we only leverage Fineweb-\ourmodel{} as training data to analysis zeroshot and data augmentation effectiveness.

\subsection{Multi-Candidates}
By now, each query is paired with the triplet of a positive document, target short answer, and target evidence bounding box.
To set up a RAG experimental environment for evaluating \ourmodel{}, 
we in addition need to let the generator take multiple candidates as input, simulating the scenario that the generator is taking multiple retrieval candidates and attributing the evidence in most relevant documents.
Given the query $q$, we use a retriever $R$ to retrieve top-$k$ candidates.
And randomly sampled $m-1$ candidates that are not ground truth as hard negative candidates.
The hard negative candidates are mixed with the one ground truth document together as the input for the multi-document \ourmodel{}.
The reason we did not directly take top-$m$ documents as the retrieval candidate is that we do not want \ourmodel{} biased on a specific retriever and position of the candidate docs.
Generally, our \ourmodel{} does not rely on the type of retriever.
It can be either a traditional text-based retriever that indexes the document with extracted text or a recent document screenshot retriever that directly indexes the original document screenshot.
However, integrating with those visual-based retrievers enables us to build an end-to-end RAG solution without the necessity of explicit document content processes such as HTML parsing or OCR.
Thus, we leverage an off-the-shelf Document Screenshot Embedding (DSE) model~\cite{ma-etal-2024-unifying} to serve as the retrieval component of the RAG system.
% For DSE, we use the off-the-shelf model available on the HuggingFace Model Hub.\footnote{https://huggingface.co/MrLight/dse-qwen2-2b-mrl-v1}
When encoding queries and documents, the model directly encodes textual queries and document screenshot images into single vector embeddings and performs cosine similarity search during inference.
In this work, we set $k=20$ and $m=3$.

Additionally, an RAG pipeline may have the chance of having no ground truth document returned from the retriever.
We use a probability of 20\% to randomly replace the ground truth document in the candidates, to access the model's capability to detect no-answer situations.
% After these crafts and filtering out queries that are not suitable for retrieval tasks, 
After these operations, the data statistics are shown in Table~\ref{tab:datasets}.

\begin{table}
\centering
\begin{tabular}{ l | c | c }
\toprule
 Dataset & \# Train  & \# Test \\ \midrule
 Wiki-\ourmodel{} & 87k & 3,000 \\  
 Paper-\ourmodel{} & 100k & 2,160  \\
 Fineweb-\ourmodel{} & 60k & - \\
\bottomrule
\end{tabular}
\caption{Datasets statistics for train and test splits.}
\label{tab:datasets}
\end{table}

\section{Experiment Setup}
\subsection{Evaluation}
Evaluation metrics assessed both the generated answers and bounding box predictions.
Relaxed exact match (EM) was used to measure generated answer accuracy, considering a generated answer correct if it shares a subsequence relationship with the golden answer and differs by no more than 20 characters. 
For bounding boxes, Intersection over Union (IoU) was calculated to determine localization precision, with an IoU threshold of 0.5 indicating a correct prediction.

To analyze performance across varying content types, test samples were categorized by the modality and location of the evidence.
For Wiki-\ourmodel{}, categories included first-page passages, passages beyond the first page, and non-passage content such as tables and figures.
For Paper-\ourmodel{}, since it is a single-page document, categories were divided into passage and non-passage content.
The overall accuracy for each dataset was computed as a macro average across these categories.

We evaluate the effectiveness of \ourmodel{} in two different settings: \textit{Single oracle candidate} and \textit{Multi-candidate}.
\textit{Single oracle candidate} setting solely evaluates the generation and visual attribution component.
We conduct controlled experiments by training and testing the VLMs using only a single ground truth relevant document screenshot as input. 
In this setup, it is guaranteed that the answer can be found within the input document.
The VLMs do not need to predict the relevant document identifier and can focus exclusively on answer generation and bounding box prediction.

In a \textit{Multi-candidate} setting, the model is evaluated on its ability to distinguish relevant documents from irrelevant ones, in addition to generating accurate answers and bounding boxes. 
This setup better reflects the RAG scenarios in which multiple candidate documents are retrieved, and the model must not only generate a correct response but also attribute it to the correct source document. 
For the \textit{Multi-candidate} evaluation, we assess two configurations:
\textit{Multi-candidate, Oracle in Candidates} which has ground truth in candidates, this setting has the same query set as the single setting, hence directly comparable.
\textit{Multi-candidate, Full} contains the additional 20\% cases where ground truth has no answer.

\subsection{Training Details}
To train vision-language models (VLMs) for answer generation with \ourmodel{}, we initialized the models using the open-source Qwen2-VL-2B and Qwen2-VL-7B~\cite{wang2024qwen2vlenhancingvisionlanguagemodels}, finetuning on the training datasets described in Section~\ref{sec:dataset}.

We first trained the models in a single-candidate setup, where the input was limited to a single oracle document image. 
In this setup, the model was trained to generate both the answer and its corresponding bounding box. We used the prompt template provided in Appendix~\ref{sec:single_prompt} to format the model's input and output.

\begin{table*}[t]
\centering
\resizebox{0.95\textwidth}{!}{%
\begin{tabular}{l|cccccccc|cccccc}
\toprule
\textbf{Method} & \multicolumn{8}{c|}{\textbf{Wiki-\ourmodel{}}} & \multicolumn{6}{c}{\textbf{Paper-\ourmodel{}}} \\
\textbf{} & \multicolumn{2}{c}{Average} & \multicolumn{2}{c}{[<1] Passage} & \multicolumn{2}{c}{[>1] Passage} & \multicolumn{2}{c|}{Non-Passage} & \multicolumn{2}{c}{Average} & \multicolumn{2}{c}{Passage} & \multicolumn{2}{c}{Non-Passage} \\
\textbf{} & bbx & ans & bbx & ans & bbx & ans & bbx & ans & bbx & ans & bbx & ans & bbx & ans \\
\midrule
\multicolumn{15}{c}{\textit{Zeroshot Prompt}} \\
% GPT4-o &  &  &  &  &  &  &  &  &  &  &  &  &  &  \\
QWen2-VL-72B & 1.5 & 60.4 & 3.4 & 58.5 & 0.1 & 54.9 & 0.9 & 67.9 & 1.5 & 43.1 & 0.5 & 40.2 & 2.5 & 45.9 \\
\midrule
\multicolumn{15}{c}{\textit{Fine-tune, Single Oracle Candidates}} \\
\ourmodel-2B-single & 37.5 & 57.1 & 70.0 & 61.1 & 18.7 & 44.9 & 23.8 & 65.3 & 63.0 & 38.3 & 50.6 & 34.4 & 75.3 & 42.1 \\
\ourmodel-7B-single & 54.2 & 65.2 & 75.6 & 66.5 & 50.1 & 56.0 & 36.8 & 73.1 & 68.2 & 43.8 & 58.1 & 41.6 & 78.2 & 45.9 \\
\multicolumn{15}{c}{\textit{Fine-tune, Multi Candidates, Oracle in Candidates}} \\
\ourmodel-2B-multi & 22.5 & 37.9 & 46.5 & 46.1 & 6.4 & 27.2 & 14.6 & 40.5 & 51.3 & 33.8 & 41.1 & 30.1 & 61.4 & 37.4 \\
\ourmodel-7B-multi & 37.7 & 41.8 & 58.1 & 49.2 & 32.8 & 32.0 & 22.2 & 44.1 & 59.9 & 39.2 & 47.7 & 35.9 & 72.0 & 42.4 \\
\multicolumn{15}{c}{\textit{Fine-tune, Multi Candidates, Full}} \\
\ourmodel-2B-full & 32.1 & 46.9 & 51.0 & 53.6 & 18.9 & 38.0 & 26.5 & 49.1 & 59.8 & 44.7 & 51.6 & 42.6 & 67.9 & 46.7 \\
\ourmodel-7B-full & 41.6 & 51.1 & 56.6 & 57.1 & 34.4 & 43.2 & 33.9 & 53.1 & 66.8 & 50.3 & 57.1 & 47.5 & 76.5 & 53.0 \\
\bottomrule
\end{tabular}%
}
\caption{Effectiveness of \ourmodel{} on Wiki-\ourmodel{} and Paper-\ourmodel{} datasets for bounding box accuracy (bbx) and answer accuracy (ans).
Fine-tuned models are trained individually on in-domain data.
The \textit{Multi-Candidate, Oracle in Candidates} setting uses the same query set as the Single Oracle Candidates setting, allowing direct comparison.
The full setting has an additional 20\% queries with no ground truth documents in candidates.}
\label{tab:main}
\end{table*}

Next, we trained the models in a multi-candidate setup. 
Here, the model received three document candidates and the task was to generate the identifier of the relevant document (if present), the answer, and the bounding box for the evidence. For cases where no relevant document was present (20\% of the training samples), the model was trained to generate ``No answer.'' We used the prompt template provided in Appendix~\ref{sec:multi_prompt} to format the model's input and output.

The training objective for both setups was next-token prediction with cross-entropy loss. We fine-tuned the models for two epochs in the single-candidate setting, using LoRA with a learning rate of 1e-4, a batch size of 64, and 4$\times$H100 GPUs. 
For the multi-candidate setting, we initialized the models with weights from the single-candidate setup and trained for one epoch with the same learning rate.
We froze the image encoder to reduce GPU memory usage in the multi-candidate setting.

During the training, random cropping was applied outside of the bounding box.
This augmentation exposed the model to varying input sizes, which enhanced its zero-shot effectiveness on unseen document layouts.
Bounding box targets were represented using absolute coordinate values.
We also explored normalizing the scale of bounding box coordinates to values in the range[0-1].
Details can be found in Section~\ref{ablation:bbx}.

\section{Experimental Results}

Table~\ref{tab:main} presents the performance of \ourmodel{} on the Wiki-\ourmodel{} and Paper-\ourmodel{} datasets across different experimental settings.
% , including zero-shot prompting and fine-tuning.
Zero-shot prompting results reveal the difficulty of directly applying state-of-the-art VLMs to the visual source attribution task.
QWen2-VL-72B achieves a reasonable answer generation accuracy of 60.4\% on average on Wiki-\ourmodel{} but fails to deliver effective bounding box predictions, with only 1.5\% accuracy.
This observation is consistent on Paper-\ourmodel{}.
% , where answer accuracy is 43.1\% but bounding box accuracy remains at 1.5\%.
These highlight the limitations of existing VLMs in pinpointing the source evidence in original documents with proper location and granularity.
% without specific task adaptation.

\begin{table*}[t]
\centering
\resizebox{0.95\textwidth}{!}{%
\begin{tabular}{l|cccccccc|cccccc}
\toprule
\textbf{Train Data} & \multicolumn{8}{c|}{\textbf{Wiki-\ourmodel{}}} & \multicolumn{6}{c}{\textbf{Paper-\ourmodel{}}} \\
\textbf{} & \multicolumn{2}{c}{Average} & \multicolumn{2}{c}{[<1] Passage} & \multicolumn{2}{c}{[>1] Passage} & \multicolumn{2}{c|}{Non-Passage} & \multicolumn{2}{c}{Average} & \multicolumn{2}{c}{Passage} & \multicolumn{2}{c}{Non-Passage} \\
\textbf{} & bbx & ans & bbx & ans & bbx & ans & bbx & ans & bbx & ans & bbx & ans & bbx & ans \\
\midrule
Wiki & 54.2 & 65.2 & 75.6 & 66.5 & 50.1 & 56.0 & 36.8 & 73.1 & 27.8 & 36.2 & 20.5 & 32.6 & 35.1 & 39.7 \\
Paper & 0.2 & 42.6 & 0 & 46.3 & 0.4 & 33.5 & 0.1 & 48.1 & 68.2 & 43.8 & 58.1 & 41.6 & 78.2 & 45.9 \\
FineWeb & 37.6 & 50.2 & 48.9 & 45.1 & 57.3 & 52.3 & 6.6 & 53.1 & 22.0 & 43.3 & 26.5 & 41.7 & 17.4 & 44.9 \\
Wiki+Fineweb & 58.2 & 65.3 & 68.7 & 66.6 & 61.7 & 57.1 & 44.1 & 72.1 & 21.0 & 43.1 & 18.5 & 42.2 & 23.4 & 43.9 \\
Paper+Fineweb & 36.1 & 48.7 & 51.8 & 49.6 & 49.6 & 44.2 & 6.8 & 52.4 & 66.5 & 44.6 & 56.1 & 42.2 & 76.9 & 47.0 \\
Wiki+Paper+Fineweb & 58.1 & 64.8 & 69.9 & 65.0 & 58.7 & 56.7 & 45.8 & 72.7 & 67.6 & 44.3 & 55.9 & 41.5 & 79.3 & 47.1 \\
\bottomrule
\end{tabular}%
}
\caption{Effectiveness of \ourmodel{} trained on different combinations training data for bounding box accuracy (bbx) and answer accuracy (ans) in the single oracle candidate setting.}
\label{tab:data}
\end{table*}

Fine-tuning on our crafted training data enables the model to effectively execute the task.
In the single-candidate setup, where the model processes only the relevant document, fine-tuned models demonstrate substantial gains compared to zero-shot prompting a much larger model.
On Wiki-\ourmodel{}, the 7B variant achieves 54.2\% bounding box accuracy and 65.2\% answer accuracy, while on Paper-\ourmodel{}, the corresponding scores reach 68.2\% and 43.8\%. 

Performance in the multi-candidate setting, which more closely mirrors real-world retrieval-augmented generation (RAG) systems, shows similar trends.
The 7B model achieves 41.6\% bounding box accuracy and 51.1\% answer accuracy when handling three candidate documents, including cases where no relevant document is present.
This demonstrates the model's capability to identify relevant sources among multiple documents while enabling fine-grained attribution.
However, when comparing the multi-candidates, oracle in candidates setting,
We can see the model facing challenges when handling multiple candidates compared to just handling a single relevant document.
E.g. on Wiki-\ourmodel{}, bounding box accuracy for 7B model is 37.7\% on average which is 17 points lower than the corresponding single candidate setting.
Showing that visual source attribution among multi-candidates is much harder than just locating the source element in a single one.
% In the full setting, where an additional 20\% ``No answer'' queries are involved, the average bounding box accuracy becomes higher (41.6\% v.s. 37.7\% for 7B model).
% We can infer that the model has high precision in telling ``No answer''. 

It further demonstrates that the effectiveness of \ourmodel{} is influenced by document characteristics, such as content location and modality.
For Wiki-\ourmodel{}, bounding box accuracy is significantly higher for passages on the first page ([<1] passage) compared to passages beyond the first page ([>1] passage).
For example, the 2B variant achieves 70.0\% accuracy for [<1] passages but only 18.7\% for [>1] passages, indicating the challenges posed by long, multi-page documents.
The larger model, the 7B variant, narrows this gap, reflecting the better handling of long-context inputs.
Non-passage content, such as tables and figures, also have obviously a different level of grounding effectiveness, indicating the difference of effectiveness in different visual elements.

% As a summary, these results underscore the effectiveness of \ourmodel{} for visual source attribution in RAG systems after fine-tuning on tailored datasets.
% However, challenges persist in achieving high bounding box precision for long-page documents or multi-candidates.

\section{Analysis}

\subsection{Out-of-Domain Zero-Shot}
Table~\ref{tab:data} shows the effectiveness of \ourmodel{} while trained with different data combinations in the single candidate setting.
It enables us to study the effectiveness of out-of-domain transfer and augmentation.
First, we highlight the challenges of zero-shot generalization in \ourmodel{}.
Training and evaluating on in-domain achieves an effective bounding box accuracy, e.g. 54.2\% on average for Wiki-\ourmodel{}.
% and training on Paper-\ourmodel{} and evaluating on Paper-\ourmodel{} achieves an even higher bbx of 68.2\%.
However, significant performance drops are observed when models are tested on out-of-domain datasets.
For instance, a model trained on Wiki-\ourmodel{} achieves only 27.8\% bounding box accuracy on Paper-\ourmodel{}, while a model trained on Paper-\ourmodel{} achieves near-zero performance (0.2\%) on Wiki-\ourmodel{}.
This gap underscores the difficulty of transferring visual source attribution capabilities across datasets with differing document structures, layouts, and content modalities.
Interestingly, Wiki-\ourmodel{} appears to transfer better to Paper-\ourmodel{} compared to the reverse.
This may be because of the multi-page nature of Wiki-\ourmodel{}, which provides richer training signals that generalize better to simpler single-page setting in Paper-\ourmodel{}.

FineWeb-\ourmodel{} shows as a promising resource for training models with improved zero-shot capabilities. When trained on FineWeb-\ourmodel{} alone, the model achieves 37.6\% bounding box accuracy on Wiki-\ourmodel{} and 22.0\% on Paper-\ourmodel{}.
Notably, FineWeb-\ourmodel{} outperforms Wiki-\ourmodel{} training on [>1] passage bbx accuracy for Wiki-\ourmodel{} (57.3\% vs. 50.1\%), suggesting its effectiveness in handling long and complex document structures. 
However, FineWeb-\ourmodel{} does not perform as well on non-passage content, likely due to its training focus on passage-level targets.

\begin{figure*}
    \centering
    \includegraphics[width=\textwidth]{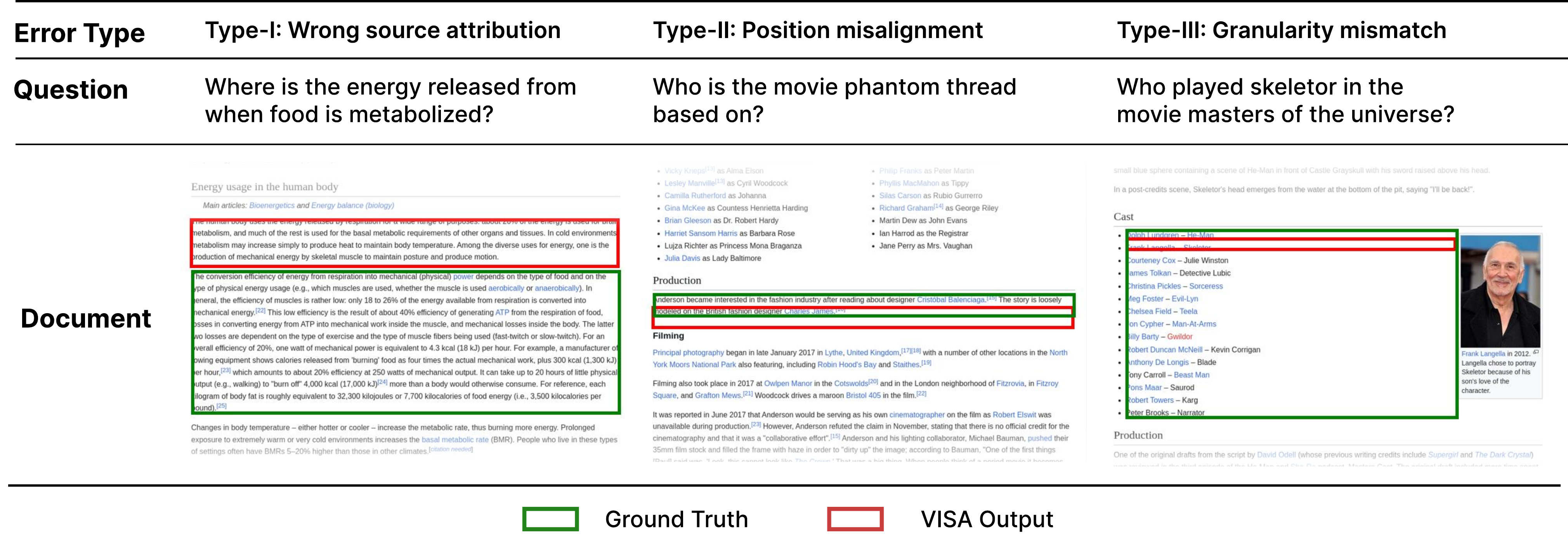}
    \caption{Type of errors in the evaluation of Wiki-\ourmodel{}.}
    \label{fig:error}
\end{figure*}
\subsection{Data Augmentation}
The results also demonstrate the benefits of augmenting training data with FineWeb-\ourmodel{}.
On Wiki-\ourmodel{}, combining Wiki and FineWeb training data improves bounding box accuracy from 54.2\% to 58.2\% and improves performance on [>1] passages from 50.1\% to 61.7\%, indicating that FineWeb complements Wiki by enhancing the model's ability to attribute evidence in multi-page contexts.
For Paper-\ourmodel{}, however, augmenting with FineWeb does not significantly improve in-domain performance.
Training on Paper+FineWeb achieves a comparable bounding box accuracy to Paper alone, but it enhances zero-shot performance on Wiki-\ourmodel{} (from 0.2\% to 36.1\%).

Training on the full combination of datasets (Wiki+Paper+FineWeb) yields strong results across both domains, with 58.1\% bbx accuracy on Wiki-\ourmodel{} and 67.6\% on Paper-\ourmodel{}. This shows the importance of diverse training data for building generalizable models capable of handling different document types, layouts, and evidence modalities. Future work should focus on expanding the dataset diversity to further improve generalization and enable robust visual source attribution for a wide range of document structures.

\begin{table}[t]
\centering
\resizebox{0.45\textwidth}{!}{%
\begin{tabular}{l|cc|cc}
\toprule
\textbf{Train Data} & \multicolumn{2}{c|}{\textbf{Wiki-\ourmodel{}}} & \multicolumn{2}{c}{\textbf{Paper-\ourmodel{}}} \\
\textbf{} & bbx & ans & bbx & ans \\
\midrule
Crop, Absolute & 54.2 & 65.2 & 27.8 & 36.2 \\
No Random Crop & 58.8 & 65.6 & 1.7 & 36.9 \\
Normalized Value & 56.4 & 64.4 & 0.1 & 37.2 \\
No Bounding Box & 0 & 67.6 & 0 & 35.2 \\
\bottomrule
\end{tabular}%
}
\caption{Impact of bounding box target representation and cropping strategies during training on Wiki-\ourmodel{} in the single oracle candidate setting.}
\label{tab:bbx}
\vspace{-0.2cm}
\end{table}

\subsection{Bounding Box Target}
\label{ablation:bbx}
Table~\ref{tab:bbx} shows the impact of different bounding box target representations and cropping strategies during training.
Training with random cropping and absolute coordinate values achieves a balance between in-domain performance on Wiki-\ourmodel{} (54.2\%) and zero-shot generalization to Paper-\ourmodel{} (27.8\%) in bounding box accuracy.
Removing random cropping slightly improves Wiki performance but drastically reduces zero-shot generalization, indicating that random cropping enhances the model’s robustness to varied input sizes.
Normalizing coordinate values achieves moderate performance on Wiki-\ourmodel{} but fails on Paper-\ourmodel{}, suggesting that absolute bounding box values are better suited to our experiments.

The ``No Bounding Box'' row represents a vanilla visual retrieval-augmented generation setup without visual source attribution, where models generate answers without bounding box predictions.
\ourmodel{} enables visual source attribution capability while the effectiveness of answer generation is preserved at about the same level of effectiveness.

\subsection{Error Analysis}
We conducted an error analysis on 50 randomly sampled cases from Wiki-\ourmodel{} to better understand the limitations of \ourmodel{}.
Errors were categorized into three main types as demonstrated in Figure~\ref{fig:error}.
The first type, wrong source attribution, occurred in 43 cases where the model attributed the source to an incorrect section of the document, failing to identify the precise region containing the evidence.
The second type, position misalignment, was observed in 4 cases where the model appeared to have the correct intent but drew the bounding box inaccurately, either slightly off position or incorrectly sized. The third type, granularity mismatch, appeared in 3 cases where the model’s attributed source, such as a specific cell in a table or an item in a list, did not match the ground truth granularity.
While these cases could potentially be considered false negatives, we leave it in error analysis to emphasize the challenge in real-world use cases where user preferences for granularity may differ from the model’s output.

\section{Conclusion}
In this paper, we introduced \ourmodel{}, a visual source attribution approach for retrieval-augmented generation pipeline. 
By leveraging vision-language models, \ourmodel{} not only generates answers to user queries but also provides bounding boxes that visually attribute the supporting evidence within document screenshots.
This capability enhances transparency and supports users in verifying the generated information effectively.
Through the development of curated datasets, we demonstrated the effectiveness of \ourmodel{} across diverse document types and layouts, including complex multi-page documents and multimodal content.
Our experimental results highlight the potential of \ourmodel{} to bridge the gap between information retrieval and answer generation by offering finer-grained, visually grounded evidence attribution.
Moving forward, we hope \ourmodel{} represents a pioneering step for more verifiable and user-friendly RAG systems.

\section{Limitations}
While \ourmodel{} demonstrates promising results for answer generation and content grounding in vision-based RAG systems, it has several limitations.
First, it focuses on generating short answers, which may not suffice for scenarios requiring detailed or explanatory responses, highlighting the need for enhancements in generating richer context.
Second, it assumes answers are derived from a single, localized region within a document, which limits its effectiveness for cases where evidence spans multiple sections or modalities (e.g., combining text and tables).
Third, while our evaluation spans web and medical scientific papers with various content modalities (e.g., passages, tables, figures), it does not fully capture the diversity of real-world documents such as scanned or handwritten content.
Additionally, as \ourmodel{} aims to make it intuitive for users to verify answers, conducting user studies could further confirm its practical utility.

\section{Acknowledgments}
We sincerely thank Xinyu Shi, Minghan Li, Jack Lin, Xinyu Zhang, and Yubo Wang for their invaluable feedback and insightful suggestions.
This research was supported in part by the Natural Sciences and Engineering Research Council (NSERC) of Canada.

\bibliography{custom}

\begin{thebibliography}{36}
\providecommand{\natexlab}[1]{#1}

\bibitem[{Asai et~al.(2024)Asai, Zhong, Chen, Koh, Zettlemoyer, Hajishirzi, and
  tau Yih}]{asai2024reliableadaptableattributablelanguage}
Akari Asai, Zexuan Zhong, Danqi Chen, Pang~Wei Koh, Luke Zettlemoyer, Hannaneh
  Hajishirzi, and Wen tau Yih. 2024.
\newblock \href {https://arxiv.org/abs/2403.03187} {Reliable, adaptable, and
  attributable language models with retrieval}.
\newblock \emph{Preprint}, arXiv:2403.03187.

\bibitem[{Bohnet et~al.(2023)Bohnet, Tran, Verga, Aharoni, Andor, Soares,
  Ciaramita, Eisenstein, Ganchev, Herzig, Hui, Kwiatkowski, Ma, Ni, Saralegui,
  Schuster, Cohen, Collins, Das, Metzler, Petrov, and
  Webster}]{bohnet2023attributedquestionansweringevaluation}
Bernd Bohnet, Vinh~Q. Tran, Pat Verga, Roee Aharoni, Daniel Andor,
  Livio~Baldini Soares, Massimiliano Ciaramita, Jacob Eisenstein, Kuzman
  Ganchev, Jonathan Herzig, Kai Hui, Tom Kwiatkowski, Ji~Ma, Jianmo Ni,
  Lierni~Sestorain Saralegui, Tal Schuster, William~W. Cohen, Michael Collins,
  Dipanjan Das, Donald Metzler, Slav Petrov, and Kellie Webster. 2023.
\newblock \href {https://arxiv.org/abs/2212.08037} {Attributed question
  answering: Evaluation and modeling for attributed large language models}.
\newblock \emph{Preprint}, arXiv:2212.08037.

\bibitem[{Carion et~al.(2020)Carion, Massa, Synnaeve, Usunier, Kirillov, and
  Zagoruyko}]{DETR}
Nicolas Carion, Francisco Massa, Gabriel Synnaeve, Nicolas Usunier, Alexander
  Kirillov, and Sergey Zagoruyko. 2020.
\newblock End-to-end object detection with transformers.
\newblock In \emph{Computer Vision -- ECCV 2020}, pages 213--229, Cham.
  Springer International Publishing.

\bibitem[{Chen et~al.(2017)Chen, Fisch, Weston, and Bordes}]{chen2017reading}
Danqi Chen, Adam Fisch, Jason Weston, and Antoine Bordes. 2017.
\newblock \href {https://aclanthology.org/P17-1171/} {Reading {W}ikipedia to
  answer open-domain questions}.
\newblock In \emph{Proceedings of the 55th Annual Meeting of the Association
  for Computational Linguistics (Volume 1: Long Papers)}, pages 1870--1879.

\bibitem[{Cheng et~al.(2024)Cheng, Sun, Chu, Xu, YanTao, Zhang, and
  Wu}]{cheng-etal-2024-seeclick}
Kanzhi Cheng, Qiushi Sun, Yougang Chu, Fangzhi Xu, Li~YanTao, Jianbing Zhang,
  and Zhiyong Wu. 2024.
\newblock \href {https://doi.org/10.18653/v1/2024.acl-long.505} {{S}ee{C}lick:
  Harnessing {GUI} grounding for advanced visual {GUI} agents}.
\newblock In \emph{Proceedings of the 62nd Annual Meeting of the Association
  for Computational Linguistics (Volume 1: Long Papers)}, pages 9313--9332,
  Bangkok, Thailand. Association for Computational Linguistics.

\bibitem[{Cho et~al.(2024)Cho, Mahata, Irsoy, He, and
  Bansal}]{cho2024m3docragmultimodalretrievalneed}
Jaemin Cho, Debanjan Mahata, Ozan Irsoy, Yujie He, and Mohit Bansal. 2024.
\newblock \href {https://arxiv.org/abs/2411.04952} {{M3DocRAG}: Multi-modal
  retrieval is what you need for multi-page multi-document understanding}.

\bibitem[{Dai et~al.(2016)Dai, Li, He, and Sun}]{NIPS2016_577ef115}
Jifeng Dai, Yi~Li, Kaiming He, and Jian Sun. 2016.
\newblock \href
  {https://proceedings.neurips.cc/paper_files/paper/2016/file/577ef1154f3240ad5b9b413aa7346a1e-Paper.pdf}
  {R-fcn: Object detection via region-based fully convolutional networks}.
\newblock In \emph{Advances in Neural Information Processing Systems},
  volume~29. Curran Associates, Inc.

\bibitem[{Dosovitskiy et~al.(2021)Dosovitskiy, Beyer, Kolesnikov, Weissenborn,
  Zhai, Unterthiner, Dehghani, Minderer, Heigold, Gelly, Uszkoreit, and
  Houlsby}]{dosovitskiy2021an}
Alexey Dosovitskiy, Lucas Beyer, Alexander Kolesnikov, Dirk Weissenborn,
  Xiaohua Zhai, Thomas Unterthiner, Mostafa Dehghani, Matthias Minderer, Georg
  Heigold, Sylvain Gelly, Jakob Uszkoreit, and Neil Houlsby. 2021.
\newblock \href {https://openreview.net/forum?id=YicbFdNTTy} {An image is worth
  16x16 words: Transformers for image recognition at scale}.
\newblock In \emph{International Conference on Learning Representations}.

\bibitem[{Faysse et~al.(2024)Faysse, Sibille, Wu, Omrani, Viaud, Hudelot, and
  Colombo}]{faysse2024colpaliefficientdocumentretrieval}
Manuel Faysse, Hugues Sibille, Tony Wu, Bilel Omrani, Gautier Viaud, Céline
  Hudelot, and Pierre Colombo. 2024.
\newblock \href {https://arxiv.org/abs/2407.01449} {Colpali: Efficient document
  retrieval with vision language models}.
\newblock \emph{Preprint}, arXiv:2407.01449.

\bibitem[{Foster(1979)}]{Foster1979}
Jeremy~J. Foster. 1979.
\newblock \href {https://doi.org/10.1007/978-1-4684-0994-9_12} {\emph{The Use
  of Visual Cues in Text}}, pages 189--203.
\newblock Springer US, Boston, MA.

\bibitem[{Gao et~al.(2023)Gao, Yen, Yu, and Chen}]{gao-etal-2023-enabling}
Tianyu Gao, Howard Yen, Jiatong Yu, and Danqi Chen. 2023.
\newblock \href {https://doi.org/10.18653/v1/2023.emnlp-main.398} {Enabling
  large language models to generate text with citations}.
\newblock In \emph{Proceedings of the 2023 Conference on Empirical Methods in
  Natural Language Processing}, pages 6465--6488, Singapore. Association for
  Computational Linguistics.

\bibitem[{Gao et~al.(2024)Gao, Xiong, Gao, Jia, Pan, Bi, Dai, Sun, Wang, and
  Wang}]{gao2024retrievalaugmentedgenerationlargelanguage}
Yunfan Gao, Yun Xiong, Xinyu Gao, Kangxiang Jia, Jinliu Pan, Yuxi Bi, Yi~Dai,
  Jiawei Sun, Meng Wang, and Haofen Wang. 2024.
\newblock \href {https://arxiv.org/abs/2312.10997} {Retrieval-augmented
  generation for large language models: A survey}.
\newblock \emph{arXiv:2312.10997}.

\bibitem[{Khalifa et~al.(2024)Khalifa, Wadden, Strubell, Lee, Wang, Beltagy,
  and Peng}]{khalifa2024sourceaware}
Muhammad Khalifa, David Wadden, Emma Strubell, Honglak Lee, Lu~Wang,
  Iz~Beltagy, and Hao Peng. 2024.
\newblock \href {https://openreview.net/forum?id=UPyWLwciYz} {Source-aware
  training enables knowledge attribution in language models}.
\newblock In \emph{First Conference on Language Modeling}.

\bibitem[{Kwiatkowski et~al.(2019)Kwiatkowski, Palomaki, Redfield, Collins,
  Parikh, Alberti, Epstein, Polosukhin, Devlin, Lee, Toutanova, Jones, Kelcey,
  Chang, Dai, Uszkoreit, Le, and Petrov}]{kwiatkowski-etal-2019-natural}
Tom Kwiatkowski, Jennimaria Palomaki, Olivia Redfield, Michael Collins, Ankur
  Parikh, Chris Alberti, Danielle Epstein, Illia Polosukhin, Jacob Devlin,
  Kenton Lee, Kristina Toutanova, Llion Jones, Matthew Kelcey, Ming-Wei Chang,
  Andrew~M. Dai, Jakob Uszkoreit, Quoc Le, and Slav Petrov. 2019.
\newblock \href {https://doi.org/10.1162/tacl_a_00276} {Natural {Q}uestions: A
  benchmark for question answering research}.
\newblock \emph{Transactions of the Association for Computational Linguistics},
  7:452--466.

\bibitem[{LeCun et~al.(2015)LeCun, Bengio, and Hinton}]{lecun2015deep}
Yann LeCun, Yoshua Bengio, and Geoffrey Hinton. 2015.
\newblock Deep learning.
\newblock \emph{nature}, 521(7553):436--444.

\bibitem[{Lewis et~al.(2020)Lewis, Perez, Piktus, Petroni, Karpukhin, Goyal,
  K\"{u}ttler, Lewis, Yih, Rockt\"{a}schel, Riedel, and Kiela}]{rag}
Patrick Lewis, Ethan Perez, Aleksandra Piktus, Fabio Petroni, Vladimir
  Karpukhin, Naman Goyal, Heinrich K\"{u}ttler, Mike Lewis, Wen-tau Yih, Tim
  Rockt\"{a}schel, Sebastian Riedel, and Douwe Kiela. 2020.
\newblock \href {https://dl.acm.org/doi/abs/10.5555/3495724.3496517}
  {Retrieval-augmented generation for knowledge-intensive nlp tasks}.
\newblock In \emph{Proceedings of the 34th International Conference on Neural
  Information Processing Systems}, NIPS '20, Red Hook, NY, USA. Curran
  Associates Inc.

\bibitem[{Lin et~al.(2024)Lin, Li, Gao, Yang, Bai, Lei, Wang, and
  Shou}]{lin2024training}
Kevin~Qinghong Lin, Linjie Li, Difei Gao, Zhengyuan Yang, Zechen Bai, Weixian
  Lei, Lijuan Wang, and Mike~Zheng Shou. 2024.
\newblock \href {https://openreview.net/forum?id=UXdxYnkJtX} {{ShowUI}: One
  vision-language-action model for generalist {GUI}}.
\newblock In \emph{NeurIPS 2024 Workshop on Open-World Agents}.

\bibitem[{Ma et~al.(2024)Ma, Lin, Li, Chen, and Lin}]{ma-etal-2024-unifying}
Xueguang Ma, Sheng-Chieh Lin, Minghan Li, Wenhu Chen, and Jimmy Lin. 2024.
\newblock \href {https://doi.org/10.18653/v1/2024.emnlp-main.373} {Unifying
  multimodal retrieval via document screenshot embedding}.
\newblock In \emph{Proceedings of the 2024 Conference on Empirical Methods in
  Natural Language Processing}, pages 6492--6505, Miami, Florida, USA.
  Association for Computational Linguistics.

\bibitem[{Malaviya et~al.(2024)Malaviya, Lee, Chen, Sieber, Yatskar, and
  Roth}]{malaviya-etal-2024-expertqa}
Chaitanya Malaviya, Subin Lee, Sihao Chen, Elizabeth Sieber, Mark Yatskar, and
  Dan Roth. 2024.
\newblock \href {https://doi.org/10.18653/v1/2024.naacl-long.167}
  {{E}xpert{QA}: Expert-curated questions and attributed answers}.
\newblock In \emph{Proceedings of the 2024 Conference of the North American
  Chapter of the Association for Computational Linguistics: Human Language
  Technologies (Volume 1: Long Papers)}, pages 3025--3045, Mexico City, Mexico.
  Association for Computational Linguistics.

\bibitem[{Min et~al.(2023)Min, Krishna, Lyu, Lewis, Yih, Koh, Iyyer,
  Zettlemoyer, and Hajishirzi}]{min-etal-2023-factscore}
Sewon Min, Kalpesh Krishna, Xinxi Lyu, Mike Lewis, Wen-tau Yih, Pang Koh, Mohit
  Iyyer, Luke Zettlemoyer, and Hannaneh Hajishirzi. 2023.
\newblock \href {https://doi.org/10.18653/v1/2023.emnlp-main.741}
  {{FA}ct{S}core: Fine-grained atomic evaluation of factual precision in long
  form text generation}.
\newblock In \emph{Proceedings of the 2023 Conference on Empirical Methods in
  Natural Language Processing}, pages 12076--12100, Singapore. Association for
  Computational Linguistics.

\bibitem[{OpenAI(2024)}]{openai2024gpt4ocard}
OpenAI. 2024.
\newblock \href {https://arxiv.org/abs/2410.21276} {{GPT-4o} system card}.
\newblock \emph{arXiv:2410.21276}.

\bibitem[{Ovadia et~al.(2024)Ovadia, Brief, Mishaeli, and
  Elisha}]{ovadia-etal-2024-fine}
Oded Ovadia, Menachem Brief, Moshik Mishaeli, and Oren Elisha. 2024.
\newblock \href {https://doi.org/10.18653/v1/2024.emnlp-main.15} {Fine-tuning
  or retrieval? comparing knowledge injection in {LLM}s}.
\newblock In \emph{Proceedings of the 2024 Conference on Empirical Methods in
  Natural Language Processing}, pages 237--250, Miami, Florida, USA.
  Association for Computational Linguistics.

\bibitem[{Penedo et~al.(2024)Penedo, Kydlíček, allal, Lozhkov, Mitchell,
  Raffel, Werra, and Wolf}]{penedo2024finewebdatasetsdecantingweb}
Guilherme Penedo, Hynek Kydlíček, Loubna~Ben allal, Anton Lozhkov, Margaret
  Mitchell, Colin Raffel, Leandro~Von Werra, and Thomas Wolf. 2024.
\newblock \href {https://arxiv.org/abs/2406.17557} {The fineweb datasets:
  Decanting the web for the finest text data at scale}.
\newblock \emph{arXiv:2406.17557}.

\bibitem[{Rashkin et~al.(2023)Rashkin, Nikolaev, Lamm, Aroyo, Collins, Das,
  Petrov, Tomar, Turc, and Reitter}]{rashkin-etal-2023-measuring}
Hannah Rashkin, Vitaly Nikolaev, Matthew Lamm, Lora Aroyo, Michael Collins,
  Dipanjan Das, Slav Petrov, Gaurav~Singh Tomar, Iulia Turc, and David Reitter.
  2023.
\newblock \href {https://doi.org/10.1162/coli_a_00486} {Measuring attribution
  in natural language generation models}.
\newblock \emph{Computational Linguistics}, 49(4):777--840.

\bibitem[{Redmon et~al.(2016)Redmon, Divvala, Girshick, and Farhadi}]{YOLO}
Joseph Redmon, Santosh Divvala, Ross Girshick, and Ali Farhadi. 2016.
\newblock \href {https://doi.org/10.1109/CVPR.2016.91} {You only look once:
  Unified, real-time object detection}.
\newblock In \emph{2016 IEEE Conference on Computer Vision and Pattern
  Recognition (CVPR)}, pages 779--788.

\bibitem[{Ren et~al.(2015)Ren, He, Girshick, and Sun}]{NIPS2015_14bfa6bb}
Shaoqing Ren, Kaiming He, Ross Girshick, and Jian Sun. 2015.
\newblock \href
  {https://proceedings.neurips.cc/paper_files/paper/2015/file/14bfa6bb14875e45bba028a21ed38046-Paper.pdf}
  {Faster r-cnn: Towards real-time object detection with region proposal
  networks}.
\newblock In \emph{Advances in Neural Information Processing Systems},
  volume~28. Curran Associates, Inc.

\bibitem[{Riedler and Langer(2024)}]{riedler2024textoptimizingragmultimodal}
Monica Riedler and Stefan Langer. 2024.
\newblock \href {https://arxiv.org/abs/2410.21943} {Beyond text: Optimizing rag
  with multimodal inputs for industrial applications}.
\newblock \emph{Preprint}, arXiv:2410.21943.

\bibitem[{Steiner et~al.(2024)Steiner, Pinto, Tschannen, Keysers, Wang, Bitton,
  Gritsenko, Minderer, Sherbondy, Long, Qin, Ingle, Bugliarello, Kazemzadeh,
  Mesnard, Alabdulmohsin, Beyer, and
  Zhai}]{steiner2024paligemma2familyversatile}
Andreas Steiner, André~Susano Pinto, Michael Tschannen, Daniel Keysers, Xiao
  Wang, Yonatan Bitton, Alexey Gritsenko, Matthias Minderer, Anthony Sherbondy,
  Shangbang Long, Siyang Qin, Reeve Ingle, Emanuele Bugliarello, Sahar
  Kazemzadeh, Thomas Mesnard, Ibrahim Alabdulmohsin, Lucas Beyer, and Xiaohua
  Zhai. 2024.
\newblock \href {https://arxiv.org/abs/2412.03555} {Paligemma 2: A family of
  versatile vlms for transfer}.
\newblock \emph{arXiv:2412.03555}.

\bibitem[{Sweller(2011)}]{SWELLER201137}
John Sweller. 2011.
\newblock \href {https://doi.org/10.1016/B978-0-12-387691-1.00002-8} {Chapter
  two - cognitive load theory}.
\newblock volume~55 of \emph{Psychology of Learning and Motivation}, pages
  37--76. Academic Press.

\bibitem[{Wang et~al.(2024)Wang, Bai, Tan, Wang, Fan, Bai, Chen, Liu, Wang, Ge,
  Fan, Dang, Du, Ren, Men, Liu, Zhou, Zhou, and
  Lin}]{wang2024qwen2vlenhancingvisionlanguagemodels}
Peng Wang, Shuai Bai, Sinan Tan, Shijie Wang, Zhihao Fan, Jinze Bai, Keqin
  Chen, Xuejing Liu, Jialin Wang, Wenbin Ge, Yang Fan, Kai Dang, Mengfei Du,
  Xuancheng Ren, Rui Men, Dayiheng Liu, Chang Zhou, Jingren Zhou, and Junyang
  Lin. 2024.
\newblock \href {https://arxiv.org/abs/2409.12191} {Qwen2-vl: Enhancing
  vision-language model's perception of the world at any resolution}.
\newblock \emph{arXiv:2409.12191}.

\bibitem[{Xia et~al.(2024)Xia, Zhu, Li, Wang, Shi, Wang, Zhang, Zou, and
  Yao}]{xia2024mmedragversatilemultimodalrag}
Peng Xia, Kangyu Zhu, Haoran Li, Tianze Wang, Weijia Shi, Sheng Wang, Linjun
  Zhang, James Zou, and Huaxiu Yao. 2024.
\newblock \href {https://arxiv.org/abs/2410.13085} {Mmed-rag: Versatile
  multimodal rag system for medical vision language models}.
\newblock \emph{Preprint}, arXiv:2410.13085.

\bibitem[{Ye et~al.(2024)Ye, Sun, Arik, and Pfister}]{ye-etal-2024-effective}
Xi~Ye, Ruoxi Sun, Sercan Arik, and Tomas Pfister. 2024.
\newblock \href {https://doi.org/10.18653/v1/2024.naacl-long.346} {Effective
  large language model adaptation for improved grounding and citation
  generation}.
\newblock In \emph{Proceedings of the 2024 Conference of the North American
  Chapter of the Association for Computational Linguistics: Human Language
  Technologies (Volume 1: Long Papers)}, pages 6237--6251, Mexico City, Mexico.
  Association for Computational Linguistics.

\bibitem[{Yu et~al.(2024)Yu, Tang, Xu, Cui, Ran, Yan, Liu, Wang, Han, Liu, and
  Sun}]{yu2024visragvisionbasedretrievalaugmentedgeneration}
Shi Yu, Chaoyue Tang, Bokai Xu, Junbo Cui, Junhao Ran, Yukun Yan, Zhenghao Liu,
  Shuo Wang, Xu~Han, Zhiyuan Liu, and Maosong Sun. 2024.
\newblock \href {https://arxiv.org/abs/2410.10594} {Visrag: Vision-based
  retrieval-augmented generation on multi-modality documents}.
\newblock \emph{Preprint}, arXiv:2410.10594.

\bibitem[{Zhao et~al.(2019)Zhao, Zheng, tao Xu, and
  Wu}]{zhao2019objectdetectiondeeplearning}
Zhong-Qiu Zhao, Peng Zheng, Shou tao Xu, and Xindong Wu. 2019.
\newblock \href {https://arxiv.org/abs/1807.05511} {Object detection with deep
  learning: A review}.
\newblock \emph{Preprint}, arXiv:1807.05511.

\bibitem[{Zhong et~al.(2019)Zhong, Tang, and Jimeno~Yepes}]{publaynet}
Xu~Zhong, Jianbin Tang, and Antonio Jimeno~Yepes. 2019.
\newblock \href {https://doi.org/10.1109/ICDAR.2019.00166} {{ PubLayNet:
  Largest Dataset Ever for Document Layout Analysis }}.
\newblock In \emph{2019 International Conference on Document Analysis and
  Recognition (ICDAR)}, pages 1015--1022, Los Alamitos, CA, USA. IEEE Computer
  Society.

\bibitem[{Zou et~al.(2023)Zou, Chen, Shi, Guo, and
  Ye}]{zou2023objectdetection20years}
Zhengxia Zou, Keyan Chen, Zhenwei Shi, Yuhong Guo, and Jieping Ye. 2023.
\newblock \href {https://arxiv.org/abs/1905.05055} {Object detection in 20
  years: A survey}.
\newblock \emph{Preprint}, arXiv:1905.05055.

\end{thebibliography}
\clearpage
\onecolumn
\appendix

\section{Appendix}

\subsection{Prompt for synthetic data generation}\label{sec:synthetic_prompt}
\begin{figure*}
    \centering
    \includegraphics[width=\textwidth]{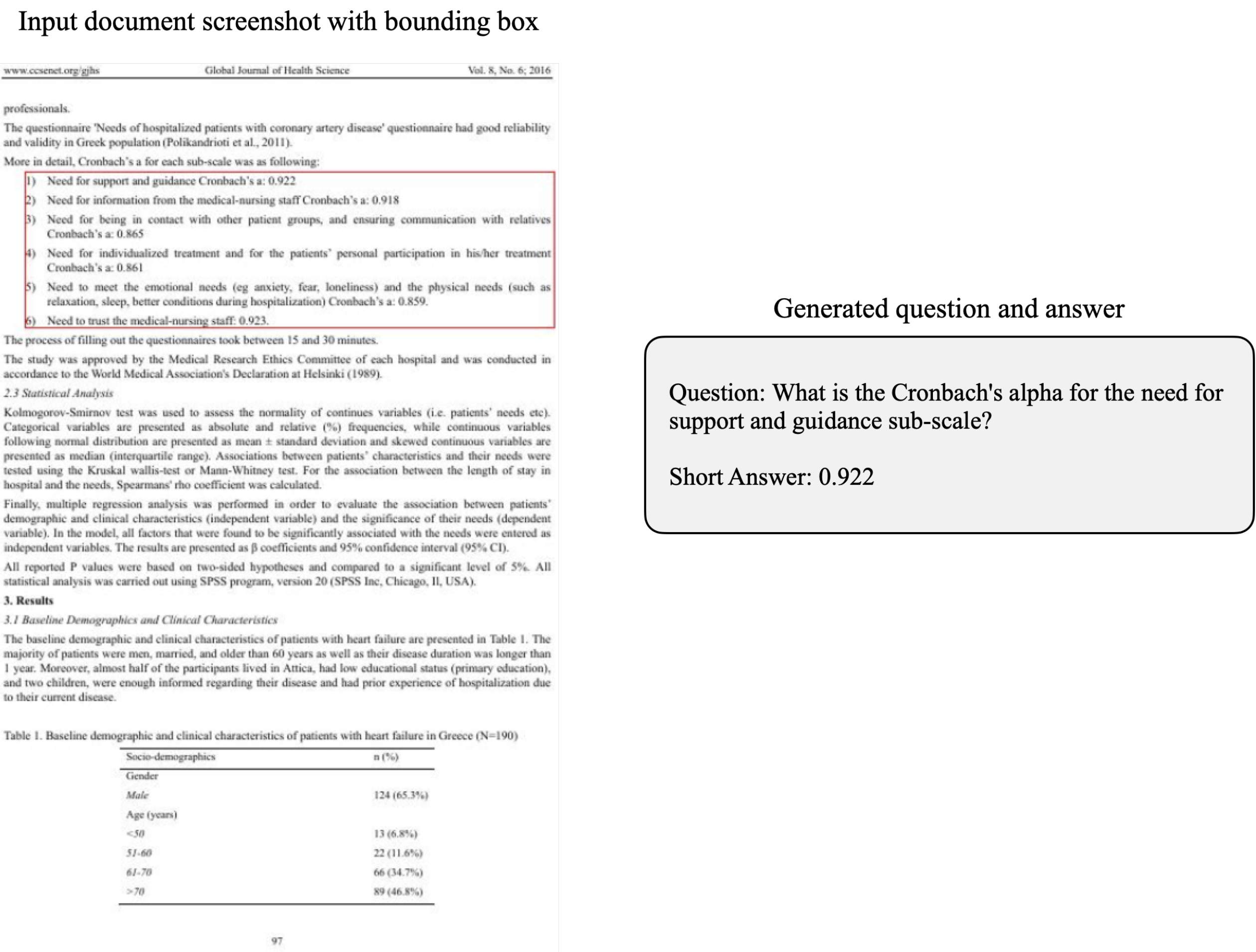}
    \caption{An example of synthetic data from Paper-\ourmodel{}.}
    \label{fig:paper-visa}
\end{figure*}

The following prompt was used for prompting QWen2-VL-72B to generate synthetic questions and answers for Paper-\ourmodel{} and Fineweb-\ourmodel{} datasets.
\begin{tcolorbox}
System:\\
Ask a question that can be specifically answered by the content in the red bounding box area and give a short answer.
The question can be a wh- question, a yes/no question, or a how question, that can be answered in a few words.\\
Output format:\\

Question: <question> \\
Short Answer: <short answer>\\

Or simply return `Empty' if the bounding box area is not visible or informative.\\

User: \{image\}
\end{tcolorbox}
\noindent Figure~\ref{fig:paper-visa} shows an example of synthetic data from Paper-\ourmodel{}.

\subsection{Prompt for Single Oracle candidate \ourmodel{}}\label{sec:single_prompt}
The following prompt template was used to format the model's inputs and outputs for training the \textit{Single Oracle Candidate} \ourmodel{}.
\begin{tcolorbox}
Model Input:\\
System:\\
Given a document image, your task is to answer the question and locate the source of the answer via a bounding box.\\

User:\\
\{image\} Image Size: \{image.size\}\\
Question: \{question\}\\

Model Output:\\
Assistant:\\
Answer: \{answer\}\\
Bounding Box: \{bounding\_box\}
\end{tcolorbox}

\subsection{Prompt for Multi-candidate \ourmodel{}}\label{sec:multi_prompt}
The following prompt template was used to format the model's inputs and outputs for training the \textit{Multi-candidate} \ourmodel{}.
\begin{tcolorbox}
Model Input:\\
System:\\
Given document images, your task is to answer the question and locate the source of the answer via a bounding box.\\

User:\\
\{image1\} Image Size: \{image1.size\}\\
\{image2\} Image Size: \{image2.size\}\\
\{image3\} Image Size: \{image3.size\}\\
Question: \{question\}\\

Model Output:\\
Assistant:\\
Answer: \{answer\}\\
Evidence Document: \{index\}\\
Bounding Box: \{bounding\_box\}
\end{tcolorbox}

\subsection{Dataset Licenses}
\begin{itemize}
    \item \textbf{NQ}: Apache License 2.0
    \item \textbf{Wikipedia}: Creative Commons Attribution Share Alike, GNU Free Documentation License family.
    \item \textbf{Fineweb-edu}: Open Data Commons License Attribution family.
    \item \textbf{PubLayNet}: Community Data License Agreement – Permissive, Version 1.0.
    
    \item \textbf{VISA Datasets}: Our crafted datasets follow the same license as the source of the documents.
\end{itemize}

\subsection{Model Backbone Licenses}
\begin{itemize}
    \item \textbf{QWen2-VL-72B}: Qwen LICENSE AGREEMENT.
    \item \textbf{QWen2-VL-2B}: Apache License.
    \item \textbf{QWen2-VL-7B}: Apache License.
    \item \textbf{VISA Models}: Our fine-tuned models follow the same licenses as the original model backbone.
    
\end{itemize}

\subsection{AI Assistant Usage}
GPT4o is used during the writing to correct grammar errors and format tables.

\end{document}